\documentclass[a4paper,fleqn]{cas-dc}

\usepackage{enumitem}
\setlist{nosep}

\usepackage[numbers,sort&compress]{natbib}
\usepackage{amsmath}
\usepackage{amssymb}
\usepackage{graphicx}
\usepackage{booktabs}
\usepackage{algorithm}
\usepackage{algorithmic}
\usepackage{listings}
\usepackage{xcolor}
\usepackage{float}

\ExplSyntaxOn
\cs_set:Npn \__first_footerline:
{
  \group_begin:
  \small
  \sffamily
  \ifnum\theblind>0\relax
  \else
    \__short_authors: :~
  \fi
  { \rmfamily \itshape Submitted~to~Data~and~Information~Management }
  \group_end:
}
\ExplSyntaxOff

\def\tsc#1{\csdef{#1}{\textsc{\lowercase{#1}}\xspace}}
\tsc{RAG}
\tsc{LLM}
\tsc{NLP}
\tsc{BM25}
\tsc{RRF}
\tsc{FAISS}

\begin{document}
\let\WriteBookmarks\relax
\def\floatpagepagefraction{1}
\def\textpagefraction{.001}

\shorttitle{Improving Contextual Retrieval for Long Documents}
\shortauthors{S. Garg et~al.}

\title [mode = title]{Improving Contextual Retrieval for Long Documents using Semantic Chunking and Hybrid Retrieval Strategies}

\author[1]{Sakshi Garg}[%
    type=editor,
    auid=1001,bioid=1]
\cormark[1]
\fnmark[1]
\ead{sakshi.garg@sjsu.edu}

\credit{Conceptualization, Methodology, Software, Data Curation, Validation, Writing - Original Draft}

\affiliation[1]{organization={Department of Computer Science, San José State University},
                addressline={One Washington Square}, 
                city={San José},
                postcode={95192}, 
                state={California},
                country={USA}}

\author[1]{William B. Andreopoulos}[%
    auid=1002,bioid=2]
\cormark[2]
\ead{william.andreopoulos@sjsu.edu}
\credit{Supervision, Project Guidance, Review \& Editing}

\cortext[cor1]{Corresponding author}
\cortext[cor2]{Principal Investigator}
\fntext[fn1]{This work was carried out as part of the MS Computer Science graduate research project (CS 298) at San José State University.}

\begin{abstract}
The vast amount of information available in lengthy, distributed documents makes finding relevant answers difficult, especially when the answer requires connecting evidence from multiple sources. Standard Retrieval-Augmented Generation (RAG) systems often struggle with this type of "multi-hop" reasoning and frequently face issues with context fragmentation. In this paper, we propose a hybrid RAG framework that integrates the structured reasoning of knowledge graphs with the broad semantic coverage of dense vector retrieval. Our approach employs a multi-stage pipeline using transformer-based summarization and semantic chunking to improve context retention. We construct a dynamic knowledge graph to capture explicit relationships between entities while simultaneously maintaining a dense vector index. During inference, the system uses a hierarchical mechanism that prioritizes graph traversal for precision and falls back to vector retrieval when necessary. We evaluated our framework on the MultiHop-RAG dataset. The results show that this hybrid approach outperforms standalone vector and graph-based baselines, achieving higher Exact Match scores and improved retrieval accuracy for complex queries. This confirms that merging structured reasoning with standard vector retrieval creates a more robust system for handling complex knowledge-intensive questions.
\end{abstract}

\begin{keywords}
Retrieval-Augmented Generation (RAG) \sep Contextual Retrieval \sep Large Language Models \sep Long Document Processing \sep Semantic Chunking \sep Hybrid Retrieval
\end{keywords}

\maketitle

\section{Introduction}

As the volume of textual information increases at an exponential rate, extracting relevant information from long texts requires innovation in Natural Language Processing (NLP) retrieval mechanisms. 
Recent developments in Large Language Models (LLMs) have transformed NLP tasks by enabling models to generate fluent, coherent, and contextually dense text. However, despite their remarkable performance, LLMs suffer from notable limitations: a lack of grounding in external knowledge and an inability to effectively manage long contexts. Retrieval-Augmented Generation (RAG) \cite{lewis2020} has emerged as a promising framework to address these shortcomings. By combining the reasoning and generation capabilities of LLMs with the contextual grounding of information retrieval systems, RAG models generate better-informed and reliable content. Document retrieval techniques along with generative models are an effective approach for optimizing long-document processing.

In traditional RAG systems, embeddings are created for documents and indexed; subsequently, the top-$k$ most relevant passages are retrieved using vector similarity search and passed as context to the model. However, these approaches often fail when the information required to answer a complex question is spread across multiple documents or when long documents contain diverse topics. RAG encounters limitations when the understanding of intricate relationships between entities is not explicitly stated within a single segment. In scenarios where semantic similarity search is imprecise, or when the user's query is ambiguous, the system may retrieve irrelevant information, hindering response accuracy. Such situations necessitate advanced mechanisms capable of multi-hop reasoning across documents to selectively extract relevant content.

To address these challenges, the LongRAG framework \cite{Zhao2024LongRAG} was designed to handle retrieval and generation in long-context settings. The authors concatenated semantically similar documents to form coherent units in a fixed-size window, leveraging advanced long-context LLMs. In our preliminary investigation, we utilized multi-hop Question Answering (QA) datasets, specifically Natural Questions (NQ) \cite{kwiatkowski2019} and HotpotQA \cite{yang2018}, to replicate the LongRAG pipeline. Our findings indicated that although long-context models (such as Gemini-1.5-Pro and GPT-4o) combined with dense retrieval performed well on benchmark metrics, these systems lacked interpretability and struggled to infer implicit relationships between entities. 

Without a structured way to traverse the logical leaps, the system remains reliant on probabilistic proximity rather than verified evidence. This limitation is best shown by two common failure modes we saw in multi-hop scenarios:

\begin{itemize}
    \item \textbf{Transitive Dependency Failures:} In cases where an answer requires linking a specific individual to a parent organization through a subsidiary, dense retrievers frequently fail to surface the intermediary document. This occurs because the intermediary chunk often lacks direct semantic overlap with the original query.

    \item \textbf{Implicit Linkage Failures:} Models often struggle to synthesize a 'hidden bridge' when the connection between two entities is implied by a shared context rather than an explicit cross-reference. While the model may retrieve the correct context, it lacks the symbolic framework to verify that these disparate facts represent a single logical chain.

\end{itemize}

These challenges motivated our pursuit of a hybrid framework that augments unstructured retrieval with symbolic reasoning.


In this paper, we propose a hybrid RAG framework that integrates a symbolic Knowledge Graph (KG), built from document content, with a traditional dense retrieval pipeline powered by a vector database. The knowledge graph encodes entities and their relationships, offering a structured representation for logical inference. When generating answers, the system can reason across paths in the KG, while dense retrieval captures the semantic nuance of the text. This combination leverages the strengths of both structured knowledge and semantic understanding to create a more accurate and reliable QA system. We evaluate this approach using the MultiHop-RAG dataset \cite{tang2024}, which represents real-world scenarios where evidence is distributed across multiple documents.

\subsection{Motivation and Problem Statement} 
Long texts introduce retrieval difficulties at several levels, including identifying topic shifts and maintaining contextual relevance within a single document. This is further compounded by information overload, particularly associated with multi-topic legal texts and scientific articles. Consequently, the ability to locate the most relevant document segments is crucial for the accurate satisfaction of user queries.

Standard RAG pipelines face two primary structural limitations. First, fixed-window chunking arbitrarily severs semantic units, leading to "context fragmentation" where essential meaning is lost across segment boundaries. Second, simply increasing the context window is ineffective due to the "lost-in-the-middle" failure mode, where models struggle to attend to information buried between the beginning and end of a long input sequence.

Furthermore, dense vector retrieval inherently flattens the hierarchical structure of complex documents. In scientific literature, semantic similarity is often insufficient; accurate retrieval requires exact terminological matches and the preservation of logical flow. A vector-only approach frequently retrieves conceptually similar but factually distinct segments. To address these failures, we propose a hybrid architecture that enforces the structural constraints of a knowledge graph alongside the semantic coverage of dense retrieval.

The remainder of this paper details the implementation and feasibility of the proposed hybrid framework. We examine relevant prior work, demonstrate the efficacy of the hybrid approach for contextual retrieval in long-document QA tasks, and discuss the potential benefits and implementation challenges of this methodology for scientific literature analysis.

\section{Background and Related Work}

Although LLMs excel at grasping human language and generating human-like text by predicting the next token, successful contextualization and retrieval from long, complex documents remain active areas of research. Work by \cite{lewis2020} highlights the capabilities of deep learning models to handle large volumes of information, such as clinical notes, to predict medical events with high precision. These developments suggest that similar techniques can be applied to improve contextual retrieval from longer texts—a challenge requiring both high-level semantic understanding and efficient retrieval mechanisms.

LLMs are transformer-based architectures that excel at generating natural language responses based on extensive training data. However, their reliance on static training sets often leads to outdated, ungrounded, or hallucinated information, particularly when addressing dynamic or confidential topics. As noted by \cite{aws2024}, LLMs face significant challenges:
\begin{itemize}
    \item \textbf{Hallucination:} They tend to generate plausible but false information when the correct answer is unknown.
    \item \textbf{Staleness:} They may present out-of-date or generic information when a specific, current response is required.
    \item \textbf{Ambiguity:} Differing sources use same terminology to describe distinct concepts, leading to terminological confusion and inaccurate responses.
\end{itemize}

LLMs are often augmented with external contextual information to mitigate these limitations. Retrieval-Augmented Generation (RAG) systems supplement LLMs with external knowledge bases—such as articles, documentation, and structured data. This augmentation expands the accessible information scope and grounds the model to generate accurate, factually relevant responses. This approach is widely employable in sectors such as healthcare, finance, and legal analysis, where domain-specific accuracy is critical.

RAG solves the problem of Large Language Models (LLMs) being static by combining parametric memory (the model's pre-trained weights) with non-parametric memory (an external, searchable index). Instead of generating responses based solely on training data, which may be outdated or incomplete, the system actively queries a reliable knowledge base during inference. This process helps by grounding the output in specific documents and mitigating the risk of hallucination in knowledge-intensive tasks.

\subsection{RAG System Architecture}
A functional RAG pipeline integrates three core subsystems:

\begin{itemize}
\item \textbf{Indexing Pipeline:} The document corpus is ingested and partitioned into discrete segments. An embedding model encodes these text chunks into dense vectors—high-dimensional numerical representations that capture semantic meaning \citep{lewis2020}.
\item \textbf{Vector Store:} A dedicated database indexes the generated embeddings to support efficient Approximate Nearest Neighbor (ANN) search. This component is responsible for retrieving context that is semantically similar to the input query.
\item \textbf{Generative Engine:} A pre-trained LLM takes the retrieved context and the original query as input to synthesize a coherent response.
\end{itemize}

This is how a Retrieval Augmented Generation pipeline works at inference time:

\begin{enumerate}
\item \textbf{Query Encoding:} The user's natural language input is mapped to the same vector space as the document index.
\item \textbf{Context Retrieval:} The system identifies top-$k$ most relevant document passages by performing a similarity search on the database. 
\item \textbf{Contextual Generation:} These passages are then concatenated with the query to form an augmented prompt. This helps LLM generate a response derived explicitly from this provided source material.
\end{enumerate}

\subsubsection{Implementation Challenges}

RAG works well in areas where factual correctness is very important, like healthcare and technical help. However, its performance is heavily dependent on the quality of the retrieval stage. Heterogeneous data formats can introduce noise, while suboptimal text segmentation (chunking) may cut off necessary semantic context. Furthermore, purely vector-based retrieval can struggle with temporal nuances. Incorporating metadata—such as document timestamps or authorship—helps to prevent the propagation of obsolete information \citep{thielsch2024}.

\subsection{Chunking Techniques in NLP}
The key to effective retrieval is coherent organization of vast amounts of unstructured text available. Chunking is the process of breaking down large documents into manageable segments called chunks while preserving original semantic meaning. This step is critical to ensure that the model processes relevant context without exceeding token limits.

Chunking strategies must balance segment size with contextual coherence. Overlapping chunks are often used to make sure that no vital information is lost at the boundaries. This is crucial for question answering and information retrieval. Table \ref{tab:chunking_comparison} summarizes common chunking strategies. \cite{jagadeesan2024}

\begin{table*}[H]
\centering
\caption{Comparison of Chunking Methods for RAG}
\label{tab:chunking_comparison}
\begin{tabular*}{\textwidth}{l @{\extracolsep{\fill}} p{0.35\textwidth} p{0.35\textwidth}}
\toprule
\textbf{Chunking Type} & \textbf{Strengths} & \textbf{Weaknesses} \\
\midrule
\textbf{Fixed-size} & Simple and fast; uniform chunk lengths & Ignores semantic structure; may split concepts \\
\textbf{Sentence-based} & Preserves sentence boundaries & May split larger coherent narratives \\
\textbf{Recursive} & Adaptive to hierarchy (paragraphs, etc.) & Complex implementation; may still break concepts \\
\textbf{Document Structure} & Preserves original formatting logic. Dependent on consistent document formatting \\
\textbf{Semantic} & High semantic fidelity; optimal for retrieval & High computational cost; dependency on embedding quality \\
\bottomrule
\end{tabular*}
\end{table*}

\subsection{Vector Databases}
Vector databases are specialized systems engineered to index and query high-dimensional floating-point vectors \cite{han2023}. In the RAG pipeline, they function as the retrieval engine for dense embeddings, employing Approximate Nearest Neighbor (ANN) methods to assess semantic similarity. In contrast to conventional inverted-index search engines (e.g., Elasticsearch), specialized vector databases like FAISS \cite{Douze2024FAISS} and Pinecone are designed for low-latency similarity searches at scale, which is essential for real-time inference.

\subsection{Knowledge Graphs}
Knowledge Graphs (KGs) systematically arrange data into a hierarchical schema comprising entities (nodes) and relationships (edges) \cite{hogan2021}. Vector search encapsulates semantic closeness, whereas knowledge graphs represent clear logical relationships. This framework enables RAG systems to execute multi-hop reasoning by navigating connections to associate facts that may be textually remote yet logically interconnected \cite{xu2024}. The main trade-off is the considerable overhead necessary for graph formation and the challenge of maintaining synchronization of the graph topology with swiftly changing underlying data.

\subsection{Graph-Enhanced and Hybrid RAG Architectures}
Hybrid architectures seek to integrate the extensive semantic scope of dense retrieval with the precision of symbolic reasoning. To overcome the limitations of vector-only retrieval, recent research has increasingly integrated Knowledge Graphs (KGs) into the RAG pipeline.

Frameworks like \textbf{HybridRAG} \cite{sarmah2024} execute concurrent retrieval procedures, employing vector search for unstructured contexts and graph traversal for structured entity associations. While this enhances accuracy in intricate domains, it often escalates system complexity by necessitating the synchronization of two separate index structures.

Recent advancements have further refined this paradigm. \textbf{GRAG} creates graph-based representations of documents to enable reasoning over structured relationships. \textbf{UniKGQA} \cite{Jiang2022UniKGQA} unifies retrieval and reasoning by distributing potential solutions along graph edges. More specifically targeting multi-hop queries, \textbf{HopRAG} \cite{tang2024} models retrieval as a "logic-aware" path-finding problem, while \textbf{SimGRAG} \cite{Cai2024} leverages subgraph similarity to improve context relevance.

Our work aligns with these approaches but differentiates itself by introducing a \textit{dynamic fallback mechanism}. Unlike purely symbolic systems that may fail when graph paths are incomplete, our system prioritizes high-precision graph traversal but seamlessly reverts to a vector-based fallback. This ensures robustness across diverse document types where structured extraction may be sparse.

\subsection{Limitations of Long-Context Architectures (LongRAG)}
An alternative to chunking and retrieval is the use of long-context LLMs (e.g., Gemini 1.5 Pro, GPT-4o) capable of processing entire documents. The \textbf{LongRAG} framework \cite{Zhao2024LongRAG} attempts to preserve global context by concatenating semantically similar documents into massive context windows (4k+ tokens).

In our preliminary reproduction of LongRAG using the Natural Questions (NQ) and HotpotQA datasets, we observed distinct failure modes despite high benchmark performance (Top-1 Retrieval Accuracy: 88\%):
\begin{itemize}
    \item \textbf{Topic Drift:} As context length increases, the model's attention mechanism often becomes diluted by irrelevant content, leading to higher inference costs and latency without proportional accuracy gains.
    \item \textbf{Segmentation Misalignment:} Simple concatenation of documents often fails to align with the specific granularity required for a query. We found that "force-feeding" long contexts does not resolve the model's inability to infer implicit relationships between distant entities.
\end{itemize}
These findings explicitly motivated our shift toward a \textit{selective hybrid} approach, where structure (KG) restricts the search space for logic, and vectors handle semantic nuance, avoiding the "lost-in-the-middle" phenomenon common in long-context models.

\section{System Methodology}

Conventional RAG implementations are generally based on flat vector retrieval. They fail to capture multi-hop dependencies in long-context scenarios. We created a hybrid architecture that integrates symbolic graph reasoning with dense vector search. 
This section delineates the system architecture, organized into three processing phases: ingestion, dual-index creation, and hierarchical inference.

\subsection{Hybrid Architecture Overview}
The proposed framework operates on a dual-path logic to balance structural precision with semantic coverage:

\begin{enumerate}
    \item \textbf{Ingestion and Preprocessing:} Raw documents are processed through a summarization model followed by a semantic chunking algorithm. This step ensures that text segments retain self-contained meaning before indexing.

    \item \textbf{Dual-Index Construction:} The processed segments fill two concurrent frameworks: a Knowledge Graph (KG) for entity-relationship mapping and a Vector Database for semantic similarity retrieval.
    
    \item \textbf{Hierarchical Inference:} The system implements a stringent priority sequence during retrieval. It initially attempts to address the query using graph traversal to identify explicit linkages. Should the graph path be incomplete or exhibit low confidence, the system initiates a fallback to a hybrid dense-sparse vector retrieval method.
\end{enumerate}

\subsection{Data Preprocessing}
The experimental corpus contains significant unstructured noise, such as redundant headers and formatting artifacts. To address this, we implemented a preprocessing pipeline that prioritizes most salient content of the document over raw text preservation.

\begin{figure}
    \centering
    \includegraphics[width=0.8\linewidth]{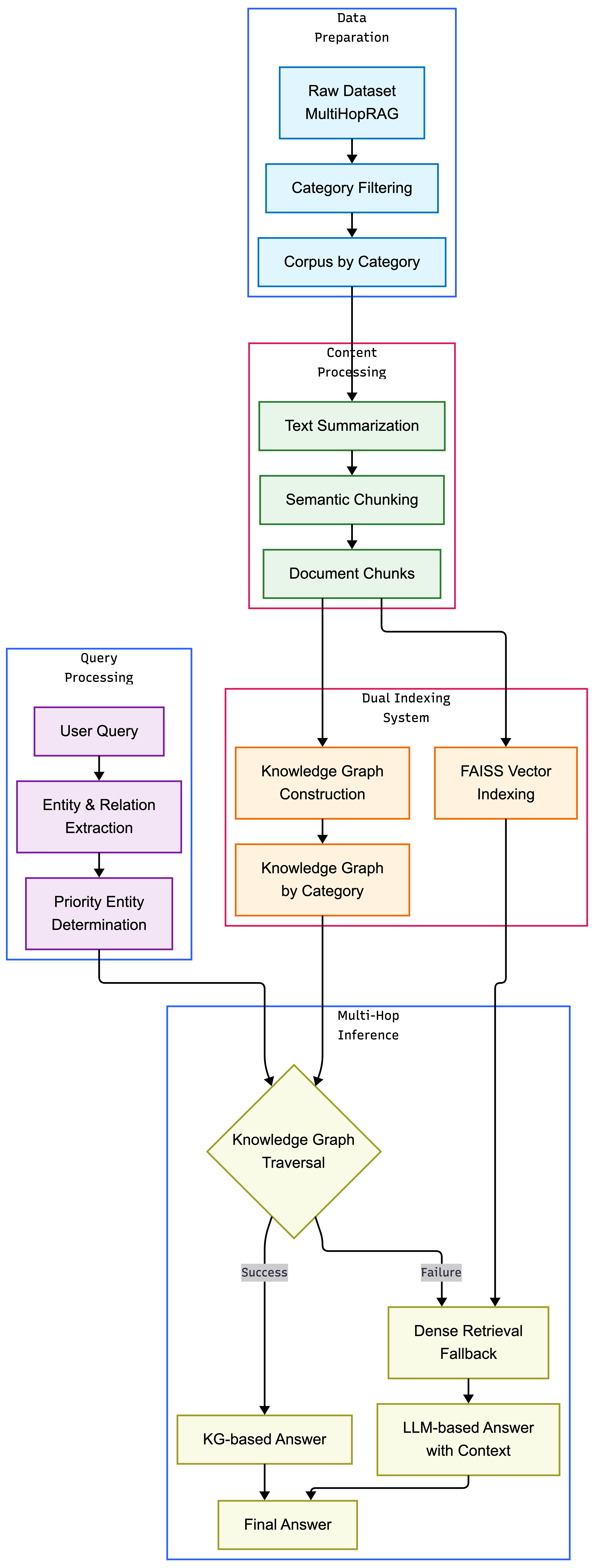}
    \caption{The end-to-end architecture of the hybrid RAG framework.}
    \label{fig:arch}
\end{figure}

\subsubsection{Document Summarization}
Initially, all the documents are passed through a summarization module based on \texttt{facebook/bart-large-cnn} transformer. This compresses the input text and helps get rid of non-essential formatting while retaining core informational content. By reducing the total token count before chunking, we decrease the computational load for downstream embedding and graph extraction tasks. The trade-off between efficiency and the retention of essential details needs careful consideration.

\subsubsection{Semantic Chunking}
Standard fixed-window chunking sometimes disrupts semantic relationships, isolating essential concepts within distinct parts. To prevent this "context fragmentation," we implemented a semantic chunking algorithm \cite{Multimodal2024}. Instead of splitting by arbitrary token counts, our method groups sentences based on embedding similarity. The algorithm computes the cosine similarity between successive sentences. Then a new segment is generated only when the similarity score falls below a dynamic threshold, signifying a change in topic. This guarantees that each segment embodies a self-sufficient concept, enhancing retrieval accuracy for cross-document inquiries.

\subsubsection{Knowledge Graph Construction and Reasoning}
The symbolic layer transforms unstructured semantic segments into a structured graph schema. Rather than depending on established ontologies, we utilize a pipeline driven by a large language model to dynamically identify entities (nodes) and relationships (edges).

\begin{figure}
    \centering
    \includegraphics[width=0.9\linewidth]{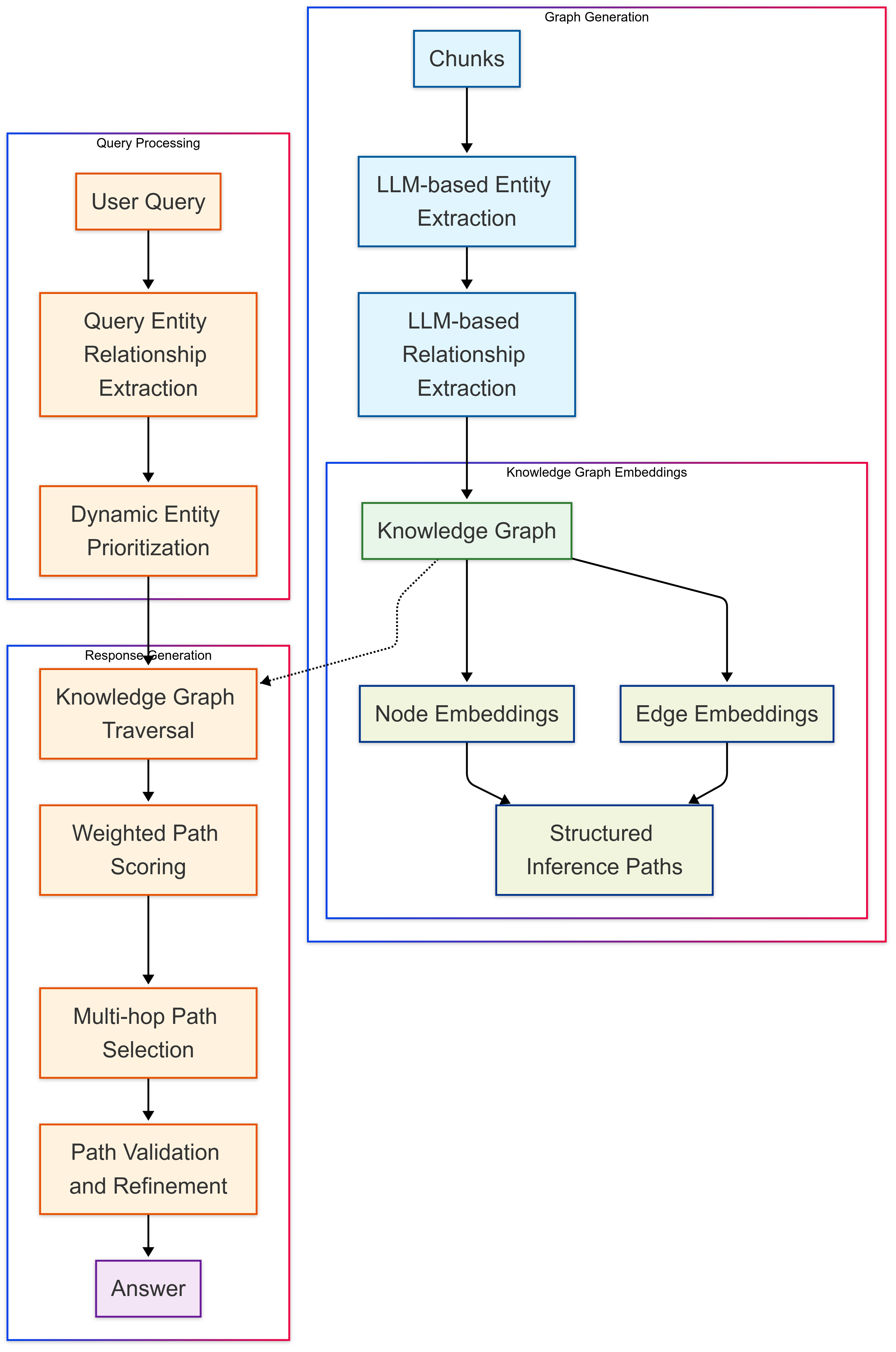}
    \caption{Knowledge Graph Reasoning}
    \label{fig:placeholder}
\end{figure}

\subsubsection{Dynamic Entity Prioritization}
To optimize graph traversal, we implement a dynamic prioritization mechanism that classifies extracted entities into four categories: \textit{subject}, \textit{key\_attribute}, \textit{related\_actor}, or \textit{source}. The system assigns an importance score $I(e)$ to each entity $e$ based on its degree centrality and relationship density within the chunk. This ensures that traversal initiates from the most information-rich nodes, reducing the search space for complex multi-hop queries.

\subsection{Dense and Hybrid Retrieval (Neural Layer)}
To ensure robustness, the system maintains a parallel retrieval pipeline. We utilize a hybrid retrieval strategy that combines dense vector embeddings with sparse keyword matching.

\subsubsection{Vector Database (Dense Retrieval)}
Semantic chunks are embedded using a pre-trained Sentence Transformer model \cite{reimers2019} and stored in a FAISS vector database \cite{Douze2024FAISS}. We utilize an IVF64-Flat index to balance retrieval speed with accuracy. This component excels at capturing abstract conceptual similarities where exact keywords may not match.

\subsubsection{Sparse Indexing and Fusion}
To complement the vector search, we maintain a sparse index (BM25) based on token frequency. This is particularly effective for queries containing specific technical terminology or proper nouns. We combine the results from the Dense (FAISS) and Sparse (BM25) indices using Reciprocal Rank Fusion (RRF) \cite{cormack2009}:

\begin{equation}
RRFscore(d) = \sum_{r \in R} \frac{1}{k + rank(d, r)}
\end{equation}

where $R$ is the set of rankings from both retrievers and $k=60$ is a smoothing constant. This fusion prioritizes documents that are consistently relevant across both retrieval paradigms.

\subsection{Path Ranking and Verification}
Once candidate paths are identified, the system employs a weighted scoring algorithm to select the most relevant reasoning chain. The score $S(p)$ for a given path $p$ is calculated as a composite of three metrics:
\begin{enumerate}
    \item \textbf{Semantic Similarity:} The cosine alignment between the path's edge labels and the user's query terms.
    \item \textbf{Evidence Count:} The frequency of the relation appearing across different document segments.
    \item \textbf{Path Depth:} A normalization factor that penalizes excessively long or circular reasoning chains to prevent hallucinations.
\end{enumerate}
This scoring mechanism ensures that the selected path balances semantic relevance with structural conciseness.

\begin{algorithm}
\caption{Hierarchical Hybrid Retrieval Inference}
\begin{algorithmic}[1]
\REQUIRE User Query $Q$, Knowledge Graph $G$, Vector Index $V$, Threshold $\tau$
\ENSURE Generated Answer $A$
\STATE \textbf{Step 1: Entity Extraction \& Prioritization}
\STATE $E \leftarrow \text{ExtractEntities}(Q)$
\STATE $E_{ranked} \leftarrow \text{Prioritize}(E)$ \COMMENT{Rank by degree centrality}
\STATE \textbf{Step 2: Symbolic Traversal}
\STATE $P \leftarrow \text{TraverseGraph}(G, E_{ranked})$ \COMMENT{Find paths connecting entities}
\STATE $P_{best} \leftarrow \text{WeightedScore}(P)$ \COMMENT{Apply composite scoring metric}
\IF{$P_{best} \neq \emptyset$ \AND $\text{Confidence}(P_{best}) > \tau$}
    \STATE $C \leftarrow \text{LinearizePaths}(P_{best})$
    \RETURN $\text{LLM\_Generate}(Q, C)$
\ELSE
    \STATE \textbf{Step 3: Vector Fallback (Self-Assessment Failure)}
    \STATE $q_{vec} \leftarrow \text{Encode}(Q)$
    \STATE $D_{dense} \leftarrow \text{Search}(V, q_{vec}, k=5)$
    \STATE $D_{sparse} \leftarrow \text{BM25}(Q, k=5)$
    \STATE $D_{final} \leftarrow \text{RRF\_Fusion}(D_{dense}, D_{sparse})$
    \RETURN $\text{LLM\_Generate}(Q, D_{final})$
\ENDIF
\end{algorithmic}
\end{algorithm}

\section{Implementation Details}

\subsection{System Configuration}
The framework is implemented in Python 3.10, utilizing specific libraries for each component of the pipeline:

\begin{itemize}
    \item \textbf{NLP Preprocessing:} \texttt{Spacy} with the \texttt{en\_core\_web\_lg} model for robust named entity recognition (NER) and tokenization.
    \item \textbf{Embeddings:} We experimented with two models:
    \begin{itemize}
        \item \texttt{all-MiniLM-L6-v2} (SentenceTransformers) for general semantic capturing.
        \item \texttt{BAAI/bge-base-en-v1.5} (FastEmbed) for high-performance dense retrieval.
    \end{itemize}
    \item \textbf{Graph Embeddings:} \texttt{Node2Vec} was used to generate 128-dimensional structural embeddings for knowledge graph nodes to facilitate graph-based similarity search.
    \item \textbf{Vector Database:} FAISS (Facebook AI Similarity Search) with IVF indexing. \cite{Douze2024FAISS}
    \item \textbf{Graph Construction:} Neo4j database populated via `LLMGraphTransformer`.
    \item \textbf{LLM Generators:} Experiments were conducted using GPT-4o and Llama-3-70B.
    \item \textbf{Compute Resources:} Local inference tests were performed on a workstation equipped with an NVIDIA RTX 3080 GPU and 32GB RAM.
\end{itemize}

\section{Experiments and Results}

In this section, we detail the dataset, evaluation metrics, and experimental design used to assess the efficacy of the proposed hybrid Retrieval-Augmented Generation (RAG) framework. We conduct a comparative analysis against two baseline architectures: (1) a standalone Dense Retrieval system (Vector RAG) and (2) a standalone Symbolic system (Graph RAG). The objective is to quantify the benefits of integrating structured symbolic reasoning with unstructured semantic search for multi-hop question answering.

\subsection{Dataset Selection}
We utilized the \textbf{MultiHop-RAG dataset} \cite{tang2024}, consisting of 2,556 human-annotated QA pairs. This dataset was selected for its high complexity, requiring reasoning across two to four distinct documents to derive a correct answer. It covers diverse domains including healthcare, business, and technology, and includes rich metadata (timestamps, authors) that mimics real-world retrieval scenarios.

For our experiments, we curated a scaled-down, balanced subset of the corpus to ensure rigorous component-level analysis. We stratified the data by domain. The dataset was processed into approximately 1,354 semantic chunks. To rigorously test the system, we generated a set of 50 complex, multi-hop queries per category (Healthcare, Sports, Technology, Business) (e.g., "How many times have the Springboks won the Rugby World Cup?") that require synthesizing information from disjoint segments. This "Custom Evaluation Set" ensures that the system is tested against real-world scenarios involving noise, formatting artifacts, and distributed evidence.

\subsection{Evaluation Metrics}
To provide a comprehensive assessment, we utilize the following metrics:
\begin{itemize}
    \item \textbf{Exact Match (EM):} The percentage of generated answers that match the gold-standard answer after normalization (lowercasing, lemmatization).
    \item \textbf{Context Precision:} Measures the proportion of relevant chunks within the retrieved context window.
    \item \textbf{Context Recall:} Measures the ability of the system to retrieve all necessary supporting facts required to answer the query.
    \item \textbf{Top-$k$ Retrieval Accuracy:} The proportion of queries where the ground-truth evidence appears within the top-$k$ retrieved segments.
    \item \textbf{Inference Latency:} The average wall-clock time required to process a query and generate a response.
    \item \textbf{KG Hit Rate:} The percentage of queries successfully answered via Knowledge Graph traversal without falling back to vector retrieval.
\end{itemize}

\subsection{Experimental Setup}
We compared the proposed Hybrid framework against two baselines:
\begin{enumerate}
    \item \textbf{Vector RAG (Baseline 1):} Utilizes standard dense retrieval via cosine similarity on text embeddings.
    \item \textbf{Graph RAG (Baseline 2):} Relies exclusively on symbolic traversal of the constructed Knowledge Graph.
\end{enumerate}

\subsection{Results}

\subsubsection{Overall Performance}
Table \ref{tab:overall_performance} presents the aggregate performance across all domains.

\begin{table}[H]
\centering
\caption{Overall Performance: Across Retrieval Strategies}
\label{tab:overall_performance}
\resizebox{\columnwidth}{!}{%
\begin{tabular}{l c c c}
\toprule
\textbf{Metric} & \textbf{Vector RAG} & \textbf{Graph RAG} & \textbf{Hybrid RAG} \\
\midrule
Exact Match (EM) & 51.5\% & 48.7\% & \textbf{55.8\%} \\
Context Precision & 66.8\% & 64.2\% & \textbf{68.1\%} \\
Context Recall & 69.9\% & 61.5\% & \textbf{72.0\%} \\
Top-5 Retrieval Accuracy & 75.5\% & 68.3\% & \textbf{78.5\%} \\
Inference Latency (s) & \textbf{1.2s} & 1.4s & 1.7s \\
KG Hit Rate & - & 64.9\% & \textbf{68.7\%} \\
\bottomrule
\end{tabular}%
}
\end{table}

The Hybrid RAG achieves the highest accuracy, demonstrating that the fallback mechanism effectively covers gaps in the Knowledge Graph. While inference latency increases slightly (1.7s vs 1.2s), the trade-off yields a significant +4.3\% improvement in Exact Match scores over the Vector baseline.

\subsubsection{Domain-Specific Performance}
We analyzed robustness across different subject matters. As shown in Table \ref{tab:domain_performance}, the hybrid approach consistently outperforms baselines, particularly in complex domains like Technology and Business where entity relationships are critical.

\begin{table}[H]
\centering
\caption{Domain-Specific Performance: Exact Match (EM)}
\label{tab:domain_performance}
\resizebox{\columnwidth}{!}{%
\begin{tabular}{l c c c}
    \toprule
    \textbf{Domain} & \textbf{Vector RAG} & \textbf{Graph RAG} & \textbf{Hybrid RAG} \\
    \midrule
    Healthcare & 49.0 & 46.5 & \textbf{53.5} \\
    Business & 52.5 & 50.2 & \textbf{55.0} \\
    Technology & 53.8 & 51.7 & \textbf{56.9} \\
    Sports & 50.1 & 47.5 & \textbf{54.3} \\
    Mixed/Other & 51.6 & 48.8 & \textbf{55.2} \\
    \bottomrule
    \end{tabular}%
    }
\end{table}

\subsubsection{Ablation Study: Preprocessing Strategies}
To validate our preprocessing pipeline, we conducted an ablation study comparing three configurations, as detailed in Table \ref{tab:preprocessing_strategies}: (1) Fixed-size chunking without summarization, (2) Fixed-size chunking with summarization, and (3) Semantic chunking with summarization (ours).

\begin{table}[H]
\centering
\caption{Preprocessing Strategies: Impact on Retrieval Quality}
\label{tab:preprocessing_strategies}
\resizebox{\columnwidth}{!}{%
\begin{tabular}{l c c c}
\toprule
\textbf{Metric} & \textbf{\shortstack{Fixed\\(No Summ.)}} & \textbf{\shortstack{Fixed\\(With Summ.)}} & \textbf{\shortstack{Semantic\\+ Summ.}} \\
\midrule
Exact Match (EM) & 39.2\% & 47.4\% & \textbf{54.9\%} \\
Context Precision & 0.43 & 0.55 & \textbf{0.68} \\
Context Recall & 0.47 & 0.61 & \textbf{0.72} \\
Top-5 Accuracy & 53.9\% & 65.4\% & \textbf{78.5\%} \\
KG Hit Rate & 49\% & 62\% & \textbf{68.7\%} \\
\bottomrule
\end{tabular}%
}
\end{table}

The data explicitly validates the "Summarize-then-Chunk" approach. Semantic chunking combined with summarization yielded a massive +16.6\% improvement in Exact Match over the naive baseline. Fixed-size chunking frequently severed semantic connections, leading to "context fragmentation."

\subsubsection{Knowledge Graph Extraction Efficacy}
The quality of the Knowledge Graph is distinctively tied to the LLM used for extraction, as shown in Table \ref{tab:llm_performance}. We evaluated three state-of-the-art models on a subset of documents.

\begin{table}[H]
\centering
\caption{Knowledge Graph Performance: Extraction across LLMs (F1-Score)}
\label{tab:llm_performance}
\resizebox{\columnwidth}{!}{%
\begin{tabular}{l c c}
\toprule
\textbf{Model} & \textbf{Entity Extr. F1} & \textbf{Relation Extr. F1} \\
\midrule
Gemini1.5-Pro (Google) & \textbf{86.5\%} & \textbf{83.2\%} \\
Llama-3-70B (Meta) & 82.3\% & 79.1\% \\
Claude 3 (Anthropic) & 84.7\% & 81.5\% \\
\bottomrule
\end{tabular}%
}
\end{table}

Google's Gemini1.5-Pro achieved the highest fidelity in extraction, likely due to its superior instruction-following capabilities for complex JSON schemas, which aided in resolving co-references across the chunk boundaries.

\subsection{Discussion}
The results substantiate the hypothesis that hybrid retrieval offers a distinct advantage for multi-hop QA.
\begin{itemize}
    \item \textbf{Synergy of Methods:} Vector RAG excels at simple, keyword-heavy queries but fails on multi-hop reasoning. Graph RAG excels at structured reasoning but suffers from sparsity (low recall) when relations are not explicitly defined. The Hybrid approach mitigates the weaknesses of both.
    \item \textbf{Latency Trade-off:} The 0.5s increase in latency is negligible for most analytical applications, justifying the precision gains.
    \item \textbf{Semantic Integrity:} The ablation study proves that \textit{how} data is ingested (Semantic Chunking) is as critical as \textit{how} it is retrieved.
\end{itemize}

\subsection{Qualitative Analysis: Inference System Trace}
To illustrate the complementary nature of the hybrid architecture, we present multiple system traces from a complex financial query involving the "FTX" dataset.

\textbf{Query 1:} "Who is the individual associated with the cryptocurrency industry facing a criminal trial on fraud and conspiracy charges and is accused by prosecutors of committing fraud for personal gain?"

\textbf{Step 1: Entity Extraction \& Prioritization}
The system extracts high-value entities: "Cryptocurrency Industry", "Criminal Trial", "Fraud", "Conspiracy". The legal charges ("Fraud", "Conspiracy") are prioritized as rigid constraints in the graph.

\textbf{Step 2: KG Traversal (Symbolic Path)}
The inference engine traverses the graph to find the node that intersects the "Founder" role with the specific "On Trial" attributes:
$$(\text{Sam Bankman-Fried}) \xrightarrow{\text{ON\_TRIAL\_FOR}} (\text{Fraud})$$
$$(\text{Sam Bankman-Fried}) \xrightarrow{\text{ON\_TRIAL\_FOR}} (\text{Conspiracy})$$
$$(\text{Sam Bankman-Fried}) \xrightarrow{\text{FOUNDER\_OF}} (\text{Cryptocurrency  \\ Exchange})$$

\textbf{Step 3: Verification}
The system verifies that the candidate node "Sam Bankman-Fried" possesses outgoing edges satisfying all three conditions: industry association (Cryptocurrency Exchange) and both legal charges.

\textbf{Generated Answer:} Sam Bankman-Fried.

\textbf{Inference Source:} Knowledge Graph.

\paragraph{\textbf{Analysis of Failure Modes:}}
While Vector RAG systems frequently retrieve documents mentioning "Sam Bankman-Fried" alongside "fraud," they often struggle to distinguish between \textit{allegations}, \textit{convictions}, and \textit{associates} facing similar charges without explicit structural grounding. A pure vector approach runs the risk of semantic drift, potentially retrieving documents about other FTX executives who appear in similar legal contexts. By anchoring the search to specific edge relations (\texttt{ON\_TRIAL\_FOR} and \texttt{FOUNDER\_OF}), the Knowledge Graph eliminates ambiguity, ensuring the answer is derived from verified logical connections rather than just keyword proximity.

\textbf{Query 2:} "Which former CTO of FTX cooperated with prosecutors and pleaded guilty to federal criminal charges at Alameda in December 2022?"

\textbf{Step 1: Entity Extraction \& Prioritization}
The system extracts key entities: "CTO", "FTX", "Alameda". Based on degree centrality, "FTX" and "Alameda" are prioritized as entry nodes for the Knowledge Graph.

\textbf{Step 2: KG Traversal (Symbolic Path)}
The system traverses the graph to find a node intersecting all constraints. The following path is discovered:
$$(\text{Gary Wang}) \xrightarrow{\text{ROLE}} (\text{Chief Technology Officer})$$
$$(\text{Gary Wang}) \xrightarrow{\text{EMPLOYER}} (\text{Alameda Research})$$
$$(\text{Gary Wang}) \xrightarrow{\text{ACTION}} (\text{Plead Guilty})$$
$$(\text{Plead Guilty}) \xrightarrow{\text{CHARGE}} (\text{Federal Criminal Charges})$$

\textbf{Step 3: Verification}
The path explicitly connects "Gary Wang" to all query constraints (Role, Company, Action).

\textbf{Generated Answer:} Gary Wang.

\paragraph{\textbf{Analysis of Failure Modes:}}
This example demonstrates how the Knowledge Graph can resolve multi-hop constraints that might be scattered across disparate document chunks. A standard Vector RAG baseline typically retrieves general articles about "FTX executives" or "Sam Bankman-Fried" because they share high semantic similarity with the query terms, but often fails to pinpoint the specific individual (Gary Wang) who meets the \textit{intersection} of all conditions (CTO + Alameda + Guilty Plea). The hybrid system succeeds because the symbolic layer enforces these structural constraints before generation.

\textbf{Query 3:} "What is the title of the book about Sam Bankman-Fried?"

\textbf{Step 1: Knowledge Graph Retrieval (Failed)}
The system attempts to query the graph for literary works linked to the entity "Sam Bankman-Fried".
\begin{itemize}
    \item \textbf{Generated KG Query:} \texttt{book about sam bankman fry}
    \item \textbf{Result:} No subgraphs found (Candidate set empty).
\end{itemize}

\textbf{Step 2: Dense Retrieval (Vector Search)}
Upon KG failure, the system triggers the vector retriever. It retrieves a high-confidence text chunk containing the answer:

 \begin{quote}
\textit{"Sam Bankman-Fried is accused of orchestrating 'one of the biggest financial frauds in American history' 'Going Infinite,' a book by Michael Lewis chronicling the rise and fall of SBF, is set to debut on the first day of the trial. Michael Lewis says he met with SBF to discuss how he could fund political candidates who were looking to offset Trump's influence in the Republican Party. Sam Bankman-Fried, the founder of failed cryptocurrency exchange FTX, is on trial for seven counts of wire fraud and conspiracy. For the first time in this trial, maybe the defense had an opening. ... The day started off promisingly for the defense as it cross-examined Gary Wang, the chief technology officer of FTX and Alameda. Bankman-Fried's fall from grace was another mark against an industry that was already roiled by bankruptcies and scandals. The government must prove intent, show that he knew what he was doing was wrong, and prove that he Lied about it anyway. ... That can potentially be embarrassing for the whole industry, Wong says."}
\end{quote}

\textbf{Step 3: Answer Extraction}
The LLM extracts the specific book title from the unstructured text context.

\textbf{Generated Answer:} Going Infinite.

\paragraph{\textbf{Analysis of Failure Modes:}}
This example highlights the "Knowledge Freshness" limitation of symbolic systems. The Knowledge Graph failed to answer because the specific entity "Going Infinite" had not yet been ingested into the static graph ontology, nor was there a predefined \texttt{BOOK\_ABOUT} relation. The hybrid architecture successfully mitigated this failure by falling back to dense retrieval. The vector component easily located the answer within the unstructured narrative of news snippets, demonstrating that dense retrieval is essential for handling dynamic information or "long-tail" queries where the graph schema is incomplete.

 \section{Conclusion}

This paper presented a novel hybrid Retrieval-Augmented Generation (RAG) framework designed to address the challenges of knowledge-intensive question answering over long and complex documents. By synthesizing the structured reasoning capabilities of Knowledge Graphs (KG) with the semantic coverage of dense vector retrieval, we successfully mitigated the specific limitations inherent to each standalone approach.

The primary contributions of this work are summarized as follows:
\begin{itemize}
    \item \textbf{Optimized Preprocessing Pipeline:} We introduced a dynamic data ingestion strategy that combines transformer-based summarization with semantic chunking. This mechanism ensures semantic coherence and reduces context fragmentation, which is critical for effective graph construction and multi-hop reasoning.
    \item \textbf{Dual-Path Inference Architecture:} We proposed a hierarchical retrieval system that prioritizes symbolic reasoning via knowledge graphs for high-precision entity traversal, while utilizing a dense vector fallback mechanism to ensure broad coverage when symbolic paths are insufficient.
    \item \textbf{Empirical Validation:} Through extensive evaluation on the MultiHop-RAG dataset, we demonstrated that the hybrid framework statistically outperforms baseline models (Vector-only and Graph-only) across key metrics, including Exact Match, Context Precision, and KG Hit Rate.
\end{itemize}

Our results indicate that while Vector RAG offers low-latency retrieval suitable for simple semantic queries, it lacks the interpretability required for complex logical deductions. Conversely, while Graph RAG provides precise reasoning paths, it suffers from sparsity issues in real-world unstructured data. The proposed hybrid RAG system effectively bridges this gap, delivering a system that is both interpretable and robust.

Furthermore, our ablation studies highlight that the quality of retrieval is heavily dependent on the upstream preprocessing steps. The integration of summarization and semantic chunking proved significantly more effective than standard fixed-window techniques. In conclusion, the proposed hybrid architecture offers a scalable and accurate solution for next-generation information retrieval systems tasked with navigating complex, multi-document corpora.

\printcredits

\section*{Declaration of Competing Interest}
The authors declare that they have no known competing financial interests or personal relationships that could have appeared to influence the work reported in this paper.

\section*{Declaration of Generative AI in Scientific Writing}
During the preparation of this manuscript, the authors used AI to assist with language editing and structural refinement. The authors reviewed and edited the content and take full responsibility for the final version of the manuscript.

\newpage
\bibliographystyle{cas-model2-names}

\begin{thebibliography}{00}


\bibitem[Lewis et al.(2020)]{lewis2020}
Lewis, P., Perez, E., Piktus, A., et al. (2020).
\newblock Retrieval-Augmented Generation for Knowledge-Intensive NLP Tasks.
\newblock \textit{Advances in Neural Information Processing Systems}, 33, 9459--9474.

\bibitem[Tang et al.(2024)]{tang2024}
Tang, Y., et al. (2024).
\newblock MultiHop-RAG: Benchmarking Retrieval-Augmented Generation for Multi-Hop Queries.
\newblock \textit{arXiv preprint arXiv:2401.15391}.

\bibitem[Zhao et al.(2024)]{Zhao2024LongRAG}
Zhao, Q., Wang, R., Cen, Y., et al. (2024).
\newblock LongRAG: A dual-perspective retrieval-augmented generation paradigm for long-context question answering.
\newblock \textit{arXiv preprint arXiv:2410.18050}.

\bibitem[Kwiatkowski et al.(2019)]{kwiatkowski2019}
Kwiatkowski, T., Palomaki, J., Redfield, O., et al. (2019).
\newblock Natural questions: A benchmark for question answering research.
\newblock \textit{Transactions of the Association for Computational Linguistics}, 7, 452--466.

\bibitem[Yang et al.(2018)]{yang2018}
Yang, Z., Qi, P., Zhang, S., et al. (2018).
\newblock HotpotQA: A dataset for diverse, explainable multi-hop question answering.
\newblock \textit{arXiv preprint arXiv:1809.09600}.

\bibitem[Sarmah et al.(2024)]{sarmah2024}
Sarmah, B., et al. (2024).
\newblock HybridRAG: Integrating Knowledge Graphs and Vector Retrieval.
\newblock \textit{arXiv preprint arXiv:2408.04948}.

\bibitem[Hogan et al.(2021)]{hogan2021}
Hogan, A., Blomqvist, E., Cochez, M., et al. (2021).
\newblock Knowledge Graphs.
\newblock \textit{ACM Computing Surveys (CSUR)}, 54(4), 1--37.

\bibitem[Douze et al.(2024)]{Douze2024FAISS}
Douze, M., Guzhva, A., Deng, C., et al. (2024).
\newblock The FAISS library.
\newblock \textit{arXiv preprint arXiv:2401.08281}.

\bibitem[Multimodal(2024)]{Multimodal2024}
Multimodal.dev. (2024).
\newblock Semantic chunking for RAG: Better context, better results.
\newblock Available at: https://www.multimodal.dev/post/semantic-chunking-for-rag [Accessed: 2025-04-19].

\bibitem[Reimers and Gurevych(2019)]{reimers2019}
Reimers, N., \& Gurevych, I. (2019).
\newblock Sentence-BERT: Sentence Embeddings using Siamese BERT-Networks.
\newblock \textit{Proceedings of EMNLP-IJCNLP}, 3982--3992.

\bibitem[Gao et al.(2022)]{Gao2022}
Gao, L., Ma, X., Lin, J., \& Callan, J. (2022).
\newblock Tevatron: An efficient and flexible toolkit for dense retrieval.
\newblock \textit{arXiv preprint arXiv:2203.05765}.

\bibitem[AWS(2024)]{aws2024}
Amazon Web Services. (2024).
\newblock Challenges in Generative AI: Hallucinations and Grounding.
\newblock \textit{AWS AI/ML Technical Report}.

\bibitem[Han et al.(2023)]{han2023}
Han, J., Pei, J., \& Tong, H. (2023).
\newblock \textit{Data Mining: Concepts and Techniques}.
\newblock Morgan Kaufmann.

\bibitem[Xu et al.(2024)]{xu2024}
Xu, Y., et al. (2024).
\newblock Retrieval-Augmented Generation with Knowledge Graphs.
\newblock \textit{arXiv preprint arXiv:2402.12345}.

\bibitem[Thielsch et al.(2024)]{thielsch2024}
Thielsch, J., et al. (2024).
\newblock The Role of Metadata in Retrieval-Augmented Generation.
\newblock \textit{Proceedings of the 2024 Conference on Data Management}.

\bibitem[Cai et al.(2024)]{Cai2024}
Cai, Y., Guo, Z., Pei, Y., et al. (2024).
\newblock Simgrag: Leveraging similar subgraphs for knowledge graphs driven retrieval-augmented generation.
\newblock \textit{arXiv preprint arXiv:2412.15272}.

\bibitem[Liu et al.(2025)]{Liu2025HopRAG}
Liu, H., Wang, Z., Chen, X., Li, Z., Xiong, F., Yu, Q., \& Zhang, W. (2025).
\newblock HopRAG: Multi-hop reasoning for logic-aware retrieval-augmented generation.
\newblock \textit{arXiv preprint arXiv:2502.12442}.

\bibitem[Jiang et al.(2022)]{Jiang2022UniKGQA}
Jiang, J., Zhou, K., Zhao, W. X., \& Wen, J.-R. (2022).
\newblock UniKGQA: Unified retrieval and reasoning for solving multi-hop question answering over knowledge graph.
\newblock \textit{arXiv preprint arXiv:2212.00959}.

\bibitem[Jagadeesan(2024)]{jagadeesan2024}
Jagadeesan, G. (2024).
\newblock Understanding chunking algorithms and overlapping techniques in natural language processing.
\newblock Available at: https://medium.com/@jagadeesan.ganesh/understanding-chunking-algorithms-and-overlapping-techniques-in-natural-language-processing-df7b2c7183b2/ [Accessed: 2025-04-18].

\bibitem[Cormack et al.(2009)]{cormack2009}
Cormack, G. V., Clarke, C. L., \& Buettcher, S. (2009).
\newblock Reciprocal rank fusion outperforms condorcet and individual rank-learning methods.
\newblock \textit{Proceedings of the 32nd International ACM SIGIR Conference}, 403--410.
\end{thebibliography}

\end{document}